# Theoretical Limits on Errors and Acquisition Rates in Localizing Switchable Fluorophores

Alexander R. Small*
Department of Physics, California State Polytechnic University-Pomona, Pomona, California

ABSTRACT   A variety of recent imaging techniques are able to beat the diffraction limit in fluorescence microcopy by activating and localizing subsets of the fluorescent molecules in the specimen, and repeating this process until all of the molecules have been imaged. In these techniques there is a tradeoff between speed (activating more molecules per imaging cycle) and error rates (activating more molecules risks producing overlapping images that hide information on molecular positions), and so intelligent image processing approaches are needed to identify and reject overlapping images. We introduce here a formalism for defining error rates, derive a general relationship between error rates, image acquisition rates, and the performance characteristics of the image processing algorithms, and show that there is a minimum acquisition time irrespective of algorithm performance. We also consider algorithms that can infer molecular positions from images of overlapping blurs, and derive the dependence of the minimum acquisition time on algorithm performance.

Recent work has shown that the diffraction limit to the resolution of an optical microscope, believed to be ~$\lambda/2$ since the work of Abbe (1), can be overcome by a variety of techniques (2). One promising road to single-molecule imaging in the far-field is to work with fluorophores that can switch between a fluorescent (activated) state and a nonfluorescent (dark) state, e.g., quantum dots (3,4), photoactivatable fluorescent proteins (5–7), or pairs of cyanine dyes (8,9). If the average distance between activated fluorophores is significantly larger than $\lambda$ then nearly all of the bright spots in the image are the result of single molecules. The centers of these bright spots can, in principle, be identified with subpixel accuracy (10,11), enabling accurate localization of individual molecules. A different subset of the fluorophores is then selected at random (via the stochastic nature of light absorption) and activated, and the process is repeated to localize the new set of activated fluorophores. Depending on the implementation, the minimum resolvable feature size ranges from 20 to 40 nm in the lateral direction and ~50 nm in the axial direction (12,13).

A common issue in all of these techniques is the number of molecules that can be activated at any one time. If too many molecules are simultaneously activated, then there is a high probability that a single bright spot will include light from multiple activated molecules and the center identified will not be an actual molecular position, introducing artifacts between two closely spaced molecules. However, decreasing the fraction of activated molecules increases the number of activation cycles needed to reliably image every molecule. Most imaging approaches with switchable molecules therefore use an algorithm to identify whether a bright spot contains one or more molecules. Typical algorithms include tests of a spot's shape (ellipticity) (9), intensity (3,14,15), or fitting to the imaging system's point spread function (6) via the CLEAN algorithm (16). If the algorithm finds that a spot contains multiple molecules, it is rejected and its center is not determined.

The questions that we address here are: Given the performance characteristics of some algorithm for rejecting bright spots with multiple molecules, what are i), the maximum possible error rate and ii), the minimum number of activation cycles needed when operating at a given error rate? Also, if instead of rejecting bright spots with small numbers of molecules we are able to solve the inverse problem and identify molecular positions, what is iii), the minimum number of activation cycles needed and how does it depend on the number of molecules the algorithm can simultaneously localize in a given bright spot?

The concept of a rejection algorithm is illustrated in Fig. 1: a spot of width $\lambda$ contains $n$ molecules. At reported resolutions of $\lambda/10$ or better, $n$ could exceed 100 in 2D or 1000 in 3D imaging. If the spot contains one activated molecule, the algorithm should accept the spot for analysis, and if the spot contains $m \geq 2$ molecules the algorithm should reject the spot. In practice, the algorithm will not work perfectly and will sometimes reject a single-molecule image or accept an $m$-molecule image. We will denote by $f_m$ ($m \geq 1$) the probability that a $m$-molecule image is accepted by the algorithm. We assume that $f_m$ is calculated by averaging over all possible positions and orientations of $m$ molecules, the distribution of light emission rates under the imaging conditions, and noise in the imaging system. The performance of the algorithm for a given imaging system and fluorescent species is completely determined by $\{f_m\}$, and an ideal algorithm has $f_1 = 1$ and $f_m = 0$ for $m \geq 2$.

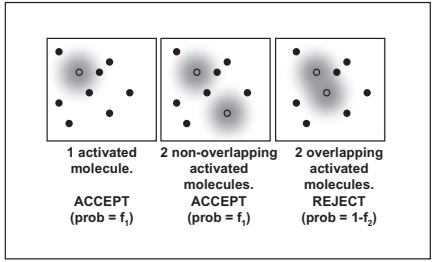

FIGURE 1 Concept of a rejection algorithm. Open and solid circles represent activated and dark molecules. When diffraction-blurred images of two molecules don't overlap, or when only one molecule is activated, the image is accepted with probability $f_1$. When blurs from two activated molecules overlap, the algorithm rejects the image with probability $f_2$.

It is straightforward to set a lower bound on the number of activation cycles needed. To ensure that almost all molecules are imaged at least once and that the variance of the number of times imaged is small (for precision in brightness measurements) we must image each molecule an average of $T \gg 1$ times; because position measurements are noise-limited, $T$ depends in part on the number of photons captured per activation cycle (10). We denote by $p$ the probability that any individual molecule is activated in a given cycle, and $p_1 = np(1-p)^{n-1}$ is the probability that a given bright spot has exactly one activated molecule. Note that the value of $p$ is determined by the dosage of the activation pulse. Over $N$ activation cycles, the expected number of times that our algorithm identifies a bright spot containing exactly one molecule will be $f_1 p_1 N = f_1 n p (1-p)^{n-1} N$, which must equal $nT$ to image each probe the desired number of times. Minimizing $N$ with respect to $p$ gives

$$p_{\text{min time}} = 1/n. \quad (1)$$

Setting $p > 1/n$ increases the acquisition time while also increasing the probability of multiple activated molecules per bright spot, so the minimum value of $N$ is $N_{\min} = nT/f_1(1 - 1/n)^{n-1}$. In the limit $n \gg 1$ (valid for high resolution), $(1 - 1/n)^n \approx e^{-1}$ (where $e$ is Euler's constant) for large $n$, giving:

$$N \geq N_{\min} = enT/f_1. \quad (2)$$

Given the value of $p$ chosen by an experimenter, we can define an error rate $E$ as the relative number of images that the algorithm accepts erroneously:

$$E = \left(\sum_{m=2}^{n} f_m p_m\right)/f_1 p_1, \quad (3)$$

where $p_m = [n!/m!(n-m)!]p^m(1-p)^{n-m}$ is the probability of activating m molecules in a given bright spot.

We will only consider errors involving two activated molecules per bright spot, and truncate the series in the numerator after the $m = 2$ term. This decision is justified in part by the fact that $p \leq 1/n$ in any efficient experiment, so $p_m/p_{m-1} \leq (n-m+1)/(n-1)m$. Also, as the number of activated molecules in a bright spot increases, the size and brightness of the spot increases, making the presence of multiple molecules easier to ascertain with good algorithms. It thus follows that $f_m$ and $p_m$ in the denominator of Eq. 3 are both decreasing functions of $m$, justifying our second order approximation. Working to second order and using $p \leq 1/n$, we therefore get an upper bound on the error rate $E$ when working with a rejection algorithm with a given $f_1$ and $f_2$:

$$E \leq E_{\max} = f_2/2f_1. \quad (4)$$

As an example, consider tests of ellipticity (9). The probability $f_2$ of accepting a two molecule image depends on the minimum molecular separation $d$ for rejection, scaling as $(d/\lambda)^2$. If $d \approx \lambda/4$ (for a blur with a significant aspect ratio), then $f_2 \approx 0.25^2 = 0.0625$ and $E_{\max} \approx 0.03125$ (for $f_1 \approx 1$). By contrast, with no rejection algorithm ($f_2 = 1$), $E_{\max} = 0.5$.

We can also derive a relationship between $N$, $E$, $f_1$, and $f_2$ by solving Eq. 3 for $p$ in terms of $E$, $f_1$, and $f_2$. in the second order approximation, and substituting that value for $p$ into $N = nT/f_1 p_1 = T/f_1 p(1-p)^{n-1}$. The result for $N$ is

$$N = \left(f_2 \frac{n-1}{2f_1 E} + 1\right) e^{2f_1 E/f_2} T. \quad (5)$$

For a low error rate $E$, the activation probability $p$ is low, and so the number of activation cycles needed to sample each molecule an average of $T$ times (not counting images where it overlaps another molecule) is inversely proportional to $E$ and directly proportional to $f_2$. However, as $2f_1 E/2f_2$ approaches 1, speed improvements are not proportional to changes in $f_2$ (Fig.2).

Our key results in Eqs. 4 and 5 are very general, applying to any algorithm that analyzes a single image of a bright spot and either accepts or rejects it based on an estimate of the number of activated molecules in the spot. One could also accept or reject a bright spot by analyzing a series of bright spots and rejecting those that occur rarely and at the center point between two more frequently occurring bright spots (based on the assumption that the rare spots are two molecule images). However, to perform that analysis one still needs to set $p$ so that the frequency of two molecule spots relative to one molecule spots does not exceed some target ratio $E$, and hence our analysis still applies.

Rejecting spots with multiple molecules is not the only possibility. Let us suppose that we have an algorithm that can solve an inverse problem for sufficiently low noise levels (e.g., a sufficiently advanced maximum likelihood estimator(17,18)) and determine molecular positions if the number of molecules present in the spot is less than or equal to some cutoff $m$. We assume that bright spots with more than $m$ molecules are either rejected or else analyzed but give unreliable results. We ask what the minimum number of activation cycles is for an ideal algorithm that can infer molecular positions in any bright spot with at most $m$ molecules ($f_r = 1$ for $r \leq m$).

If we wish to accurately (i.e., in a spot with $m$ or fewer molecules) image all $n$ molecules an average of $T$ times, the number of activation cycles $N$ is given by

$$nT = N \sum_{r=1}^{m} r p_r$$
$$= N \sum_{r=1}^{m} \frac{n(n-1)\ldots(n-m+1)}{(r-1)!} p^r (1-p)^{n-m}. \quad (6)$$

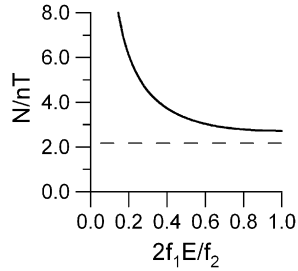

FIGURE 2  Relationship between number of activation cycles required for a given error rate $E$ and rejection algorithm performance characteristics $f_1$ and $f_2$.

Minimizing $N$ with respect to $p$ gives

$$\sum_{r=0}^{m-1} \frac{(n-1)\ldots(n-r)}{r!} p^r (1-p)^{n-r-1}$$
$$= \frac{(n-1)\ldots(n-m)}{(m-1)!} p^m (1-p)^{n-1-m}. \quad (7)$$

We assume an approximate solution $p = a/n$ where $a$ is a constant to be determined. For $n \gg r$, we can make the approximations $(n-1)\ldots(n-r) \approx n^r$ and $(1-a/n)^n \approx e^{-a}$ and get

$$\sum_{r=0}^{m-1} a^r/r! = a^m/(m-1)! \quad (8)$$

The solution is $a = 1$ for $m = 1$ and $a = (1+\sqrt{5})/2 = 1.62$ for $m = 2$. For $m > 2$, Eq. 8 can be easily solved numerically and solutions for small $m$ are given in Table 1. In the limit of large $m$, $a \to m$. When we use $p = a/n$ and our values for $a$ in Eq. 6, we get the values for $N$ shown in Table 1. Notice that $N$ decreases more slowly as $m$ increases. The asymptotic form of $N$ for $n \gg m \gg 1$ is $N \propto nT/m$.

In conclusion, we have derived a formula relating the maximum achievable image acquisition rate and the error rate in imaging of switchable fluorophores, relating these to the performance of the algorithm being used to reject bright spots with multiple fluorophores. For low error rates, when the average number of fluorophores activated well below one per bright spot, the acquisition timescales inversely with the error rate. For large error rates, the acquisition rate decreases more slowly than $1/E$, reaching a minimum when an average of one molecule per bright spot is activated. For any given rejection algorithm there is a maximum acquisition error rate. Algorithms that solve an inverse problem for bright spots with multiple activated fluorophores and identify positions of molecules can enable acquisition rates faster than those achievable with rejection algorithms, scaling inversely with the number of molecules localized per bright spot.

TABLE 1  Optimal activation probabilities and activation cycles

| $m$ | $a = p \times n$ | $N/nT$ |
| --- | --- | --- |
| 1 | 1 | $e = 2.71828$ |
| 2 | 1.62 | 1.19 |
| 3 | 2.27 | 0.73 |
| 4 | 2.94 | 0.51 |
| 5 | 3.64 | 0.39 |
| 6 | 4.35 | 0.32 |